\documentstyle[preprint,aps]{revtex}
\widetext
\draft
\tighten

\begin{document}

\title{The phase free, longitudinal, magnetic component of vacuum
electromagnetism}

\bigskip

\author{{\bf A.E. Chubykalo},  {\bf M.W. Evans}\thanks{Department of Physics 
and Astronomy, York University,  North York, Toronto, Canada} and 
{\bf R. Smirnov-Rueda}
\thanks{Instituto de Ciencia de Materiales, C.S.I.C., Madrid, Spain}}

\address {Escuela de F\'{\i}sica, Universidad Aut\'onoma de Zacatecas \\
Apartado Postal C-580\, Zacatecas 98068, ZAC., M\'exico}

\date{\today}

\maketitle

%  \date{December 12, 1995}

\baselineskip 7mm

\begin{abstract}
       A charge $q$ moving in a reference laboratory system with constant
velocity {\bf V} in the $X$-axis produces in the $Z$-axis a longitudinal, 
phase free, vacuum magnetic
field which is identified as the radiated ${\bf B}^{(3)}$ field of Evans, 
Vigier and
others.

\end{abstract}

\pacs{PACS numbers: 03.50.De, 03.50.Kk}

\newpage
       Several inferences have converged recently on the renewed
conclusion that vacuum electromagnetism is three dimensional, not
transverse as in the received view. The problem of longitudinal
electromagnetic field components in vacuo has been put forward
by Majorana [1], Dirac [2], Oppenheimer [3] and Wigner [4],
who inferred a discrete phase free variable, the spin. Much later,
``acausal" fields of this type were given independently by Gianetto [5]
and by Ahluwalia and Ernst [6].  The relativistic, three dimensional
soliton theory of Hunter and Wadlinger [7] implies the same conclusion,
supported empirically.  Other empirically supported theories that give
longitudinal fields in vacuo include those of Recami et al. [8] and
Rodrigues et al.  [9].  Meszaros et al. [10] have produced a
thermodynamically based theory leading to the same result, whose
ramifications have also been developed by Lehnardt [11]. Dvoeglazov [12]
has reviewed circa 150 papers which infer non-Maxwellian properties in
vacuo. Dvoeglazov et al. [13] have discussed inconsistencies between the
Joos-Weinberg and Maxwell equations.  A substantial work by Chubykalo and
Smirnov-Rueda [14] removes several well-known inconsistencies in classical
electrodynamics by invoking simultaneously transverse and longitudinal
components in vacuo. Munera and Guzman [15] in three recent papers, have
arrived at the existence of longitudinal components and the magnetic
scalar potential using a rigorous re-examination of the Lorentz condition.
Finally, the theory of the ${\bf B}^{(3)}$ field and of the {\bf B} cyclic
equations has been presented in several recent monographs [16] which
develop the subject systematically to show that in general, longitudinal
solutions are linked to transverse counterparts by a new equivalence
principle. In this Letter it is shown that the theory of Chubykalo and
 Smirnov-Rueda [14] leads directly to the ${\bf B}^{(3)}$ field of Evans,
Vigier and others [16]. These two lines of thought converge on the same
conclusion.

           To see this, use Gaussian units and consider a charge $q$ moving
in a reference laboratory frame with a constant velocity {\bf V} along the
positive $X$-axis. Let the site of the charge at instant $t$ be  ${\bf r}_q$,
 $(x_q, 0, 0)$. Maxwell's displacement current is zero in this theory
everywhere . The law of Biot and Savart
[17] gives, for this system, the magnetic field strength:
\begin{equation}
{\bf H}=\frac{1}{c}{\bf V}\times {\bf E}
\end{equation}
where the {\bf E} is given by [19]:
\begin{equation}
{\bf E}=(1-\beta^2)\frac{q{\bf R}}{R^3(1-\beta^2\sin^2\theta)^{3/2}}
\end{equation}
were $R$ is distance between the charge and  a point of observation (in 
our case $R=\biggl[X(t)^2+y^2+z^2\biggr]^{1/2}$,  $X(t)=x-x_q(t)$.

     Using Amp\`ere's Law [17] without Maxwell's displacement current gives
$curl\,\,{\bf H}=\frac{4\pi}{c}{\bf j}$ where {\bf  j} is the conducting
current density ${\bf j} = \varrho{\bf V}$.  Use of Gauss's Theorem [17]
 $div\,\,{\bf  E} = 4\pi\varrho$ results in:
  \begin{equation}
curl\,\,{\bf H}=\frac{1}{c}{\bf V}(div\,\,{\bf E})=\frac{1}{c}curl({\bf
V}\times{\bf E})+\frac{1}{c}({\bf
V}\cdot\nabla){\bf E}
 \end{equation}
 (using $div\,\,{\bf V}=({\bf
E}\cdot\nabla){\bf V}=0$).
However,
from eqn.  (1):  \begin{equation} curl\,\,{\bf H}=\frac{1}{c}curl({\bf
V}\times{\bf E})
 \end{equation}
 and eqns. (3) and (4) produce a paradox, because
$({\bf V}\cdot\nabla){\bf E}$  is rigorously non-zero. There is a term
needed to cancel out the first term on the right hand side of eqn. (3),
which has been derived in the steady state [17] assuming that there is no
change in net charge density anywhere in space, i.e. by using the Amp\`ere's
Law without Maxwell's displacement current. The missing term must
therefore originate in an entirely {\it novel} displacement current, ${\bf
j}_d$, hitherto unconsidered in electrodynamics.  Thus Amp\`ere's Law
becomes:  \begin{equation} curl\,\,{\bf H}= \frac{4\pi}{c}({\bf j} + {\bf
      j}_d).  \end{equation} We know that \quad $div\,curl\,\,{\bf H} =
0$\quad  from vector analysis [18]; so, since ${\bf j}_d$ is not Maxwell's
 famous displacement current by construction, (thus $div\,\,{\bf j}_d =
0$), the only possible alternative is:  \begin{equation} {\bf j}_d =
          \frac{1}{4\pi}curl (\cal U{\bf F}) \end{equation} where ${\cal
U}(x, y,z, t)$ and ${\bf F}(x, y, z, t)$ are scalar and vector functions
of space and time. We also note that the solution (6) is part of a more
general, well-known, equation [17]:
$$
              div\,\,{\bf j}_d = \frac{1}{4\pi}div\Biggl(\frac{d{\bf E}}{dt}
              \Biggr).
$$
From eqn. (3), it is seen that {\bf F} is in the $Z$-axis,
mutually perpendicular to $V_x$ and $E_y$; and has been introduced
in the context of a steady state, {\it phase  free}, problem. Also, 
${\cal U}{\bf F}/c$
has the units of magnetic field strength, which we denote ${\bf H}^{(3)}$. 
This is
clearly the analogue of ${\bf B}^{(3)}$ [16].
Eqns. (3) and (4) become the same therefore if \footnote{
The rigorous derivation of eqn. (7) requires the separation of
fields [14]:
$$
         {\bf E}_{(tot)} = {\bf E}_0 + {\bf E}^*
$$
where ${\bf E}_0$ becomes the solution of Poisson's equation in the static
limit, and where ${\bf E}^*$ is the solution of the wave equation for free 
field. Therefore
${\bf E}^*$ is a function of retarded time, but ${\bf E}_0$ is not. This 
requires a
careful re-examination of precepts in partial differential analysis, and
we have carried this out in the course of our derivation of eqn. (7). More
details will be reported in future work. Eqn. (7) is rigorously correct
if and only if ${\bf E}_0$  is a function of the type
${\cal F}(X(T), y, z)$, where time $T$ does not dependent on retarded 
time ($T$ is not denoted by the retarded time); and if ${\bf E}^*$ is a
function of the type ${\cal F}(x, y, z, t)$ where $t$ is compound function 
of retarded time ($t$ is denoted by the retarded time and vice versa).}:
\begin{equation}
curl({\cal U}{\bf F})=-({\bf V}\cdot\nabla){\bf E}.
\end{equation}
In source free regions of space (i.e. very far from the charge) we
obtain:
\begin{equation}
curl({\cal U}{\bf F})=0
\end{equation}
Since {\bf F} is phase free in the vacuum, its curl is zero, and so:
\begin{equation}
grad\,\,{\cal U}\times{\bf F}=0
\end{equation}
If {\bf F} is in the $Z$-axis by construction it is given from eqn. (9),
finally, by:
\begin{equation}
F_z=-\Biggl( \frac{\partial{\cal U}}{\partial z}\Biggr)^2 w
\end{equation}
where $w$ is an arbitrary constant scalar.

{\it This is a phase free, radiated, longitudinal magnetic field, which can 
exist in the absence or presence of Maxwell's displacement current, and which 
is produced by our novel displacement current}   ${\bf j}_d$.

         Thus {\bf F} has the same properties precisely as the previously
inferred ${\bf B}^{(3)}$ magnetic flux density [16]. It is the radiated 
longitudinal
magnetic field due to the infinitely distant charge $q$. Such a field does
not exist in the received view in the absence of Maxwell's displacement
current ${\partial{\bf E}}/{\partial t}$. Furthermore, since $curl\,\,{\bf
F} = 0$ in vacuo, it follows that ${\bf F} = grad\,\,\varphi_m$, where
$\varphi_m$ is the magnetic scalar potential of Munera and Guzman [15].
Also, since $div\,\,{\bf F} = 0$ in vacuo, then ${\bf F} = curl\,\,{\bf
A}$; and so $curl\,\,{\bf A} = grad\,\,\varphi_m$ in vacuo. This leads to
the magnetic dual interpretation of Maxwell's equations by Munera and
Guzman [15], who used the conventional displacement current. In general,
${\bf B}^{(3)}$ coexists with, and is linked geometrically to, the transverse
irradiated wave component ${\bf B}^{(1)} = {\bf B}^{(2)*}$ [16] through 
the vacuum {\bf B} Cyclic
equations. The transverse irradiated waves, however, are phase dependent
in vacuo. The field {\bf F} can exist when {\bf E} (free) is not zero and 
${\bf V}=0$ because determinants of eqns. (7) and (9) are zero and
eqn. (9) must have a non-zero solution, even when all minors of (9) are zero.
In other words, this is true even when {\bf E} on the right hand side of eqn. 
(7) is zero, i.e.
when the only field present is the irradiated (source free) field. The
results of our calculation are different from those of Jackson [17],
p. 381, where the relativistic radiation from a charge translating with
constant velocity is shown to be a plane polarized transverse wave, with
an oscillating longitudinal component. Jackson uses implicitly Maxwell's
displacement current because the non-zero field components resulting from
his calculation are time dependent. A complete understanding of this
basic problem in electrodynamics requires therefore consideration of {\it both}
the Maxwell displacement current and our novel current ${\bf j}_d$. This
should produce, consistently, the {\bf B} cyclic Theorem in vacuo, i.e.
\begin{equation}
              {\bf B}^{(1)}\times{\bf B}^{(2)} = i B^{(0)}{\bf B}^{(3)*}
\end{equation}
in cyclic permutation in the basis $((1), (2), (3))$ [16].

\acknowledgments
MWE  is grateful to the York University, Toronto; and the Indian Statistical 
Institute
 for visiting professorships. Many colleagues are thanked for
e-mail discussion and preprints of related work. AECh acknowledges 
many stimulating discussions with  V. V. Dvoeglazov.


\begin{references}

 \bibitem{1} R. Mignani, E. Recami and M. Baldo, Lett. Nuovo Cim., {\bf 11},  
 568 (1974).
\bibitem{2}  P.A.M. Dirac, {\it Directions in
Physics} (Wiley, New York, 1978).
\bibitem{3} J.R. Oppenheimer, Phys. Rev., {\bf 38}, 725 (1931).
\bibitem{4} E.P. Wigner, Ann. Math., {\bf 40}, 149 (1939).
\bibitem{5} E. Gianetto, Lett. Nuovo Cim., {\bf 44}, 140 (1985).
\bibitem{6} D.V. Ahluwalia and D.J. Ernst, Mod. Phys. Lett., {\bf A7}, 1967 
(1992).
\bibitem{7} G. Hunter and R.L.P. Wadlinger, Phys. Essays, {\bf 2}, 156 (1989).
\bibitem{8} A.O. Barut, G.D. Maccarone and E. Recami, Nuovo Cim., {\bf A71}, 
509
(1982); V.S. Olkhovsky and E. Recami, Phys. Rep., {\bf 214}, 339 (1992); W.
Heitman and G. Nimtz, Phys. Lett., A, {\bf 196}, 154 (1994); E. Recami, 
Rivista Nuovo Cim., 9(6) (1986).
\bibitem{9} W.A. Rodrigues, Jr., and J.-Y. Liu, Institute of Mathematics, 
State
University of Campinas, Brazil, RP 12/96 (1996).
\bibitem{10} M. Meszaros, Found. Phys. Lett., submitted for publication; M.
Meszaros and P. Molnar, work in progress.
\bibitem{11} B. Lehnardt, in M. W. Evans, J.-P. Vigier, S. Roy and G. Hunter 
(eds.),
{\it The Enigmatic Photon,  New Developments.}, Vol. 4 (Kluwer,
Dordrecht, 1997), in preparation.
\bibitem{12} V.V. Dvoeglazov, ibid., in preparation.
\bibitem{13} V.V. Dvoeglazov, Yu.N. Tyukhtyaev and S.V.  Khudyakov, Russian J.
Phys., {\bf 37}, 898 (1994).
\bibitem{14} A.E. Chubykalo and R. Smirnov-Rueda, Phys. Rev. E, in press (May
1996); ibid., in preparation.
\bibitem{15} H. Munera and O. Guzman, Found. Phys. Lett., in press.
\bibitem{16} M.W. Evans, J.-P. Vigier, S. Roy and S. Jeffers, {\it 
The Enigmatic
Photon.} Vols 1-3 (Kluwer, Dordrecht, 1994, 1995, 1996); M.W. Evans, Physica B,
{\bf 182}, 227 (1992); Physica A, {\bf 214}, 605 (1995); Found.
Phys., 24, 892, 1519, 1671 (1994); 25, 175, 383 (1995); Found. Phys.
Lett., 7, 67, 209, 467 (1994); 8, 63, 83, 187, 363, 385 (1995); also
papers in Found. Phys. Lett., {\bf  9}, (1996).
\bibitem{17} J.D. Jackson, {\it Classical Electrodynamics.} (Wiley, New York, 
1962).
\bibitem{18} G. Stephenson, {\it Mathematical Methods for Science Students.} 
(Longmans
Green and Co., London, 1968, fifth impression).
\bibitem{19} L.D. Landau and E.M. Lifshitz, {\it Teoria Polia} (Nauka, Moscow, 
1973)  [English translation: {\it Classical Theory of Field} (Pergamon, 
Oxford, 1985)] 
\end{references}
\end{document}